\documentclass[12pt,preprint]{aastex}
\begin{document}
\title{HST and Optical Observations of Three Pulsating Accreting White
Dwarfs in Cataclysmic Variables\footnote{Based on observations made with the 
NASA/ESA Hubble Space
Telescope, obtained at the Space Telescope Science Institute, which is
operated by the Association of Universities for Research in Astronomy, Inc.,
under NASA contract NAS 5-26555, with the Apache Point Observatory 3.5m 
telescope which is owned and operated by the Astrophysical Research
Consortium, and with the Nordic Optical Telescope, operated on the island
of La Palma jointly by Denmark, Finland, Iceland, Norway and Sweden, in the
Spanish Observatorio del Roque de los Muchachos of the Instiitute de
Astrofisica de Canaries.}}
\author{Paula Szkody\altaffilmark{2},
Anjum Mukadam\altaffilmark{2},
Boris T. G\"ansicke\altaffilmark{3},
Patrick A. Woudt\altaffilmark{4},
Jan-Erik Solheim\altaffilmark{5},
Atsuko Nitta\altaffilmark{6},
Edward M. Sion\altaffilmark{7},
Brian Warner\altaffilmark{4},
D. K. Sahu\altaffilmark{8},
T. Prabhu\altaffilmark{8},
Arne Henden\altaffilmark{9}}
\altaffiltext{2}{Department of Astronomy, University of Washington, Box 351580,
Seattle, WA 98195}
\altaffiltext{3}{Dept. of Physics, University of Warwick, Coventry CV4 7AL, UK}
\altaffiltext{4}{Dept. of Astronomy, University of Cape Town, Private Bag, Rondebosch 7700, South Africa}
\altaffiltext{5}{Institute of Theoretical Astrophysics, Box 1029 Blindern, N-0315 Oslo, Norway}
\altaffiltext{6}{Gemini Observatory, 670 N A'ohoku Pl., Hilo, HI 96720 and
Subaru Telescope, 650 N A'ohoku Pl., Hilo, HI 96720}
\altaffiltext{7}{Dept. of Astronomy \& Astrophysics, Villanova University, Villanova, PA 19085}
\altaffiltext{8}{Indian Institute of Astrophysics, Koramangala, Bangalore 560 034, India}	
\altaffiltext{9}{American Association of Variable Star Observers, 25 Birch Street, Cambridge, MA 02138}

\begin{abstract}
Ultraviolet observations using the Solar Blind Channel on the Hubble
Space Telescope provide light curves and low resolution spectra of
three pulsating white dwarfs in the cataclysmic variables 
SDSS013132.39-090122.3, SDSSJ161033.64-010223.3 and SDSSJ220553.98+115553.7. 
The UV light curves show
enhanced pulsation amplitudes over those from simultaneous and previous optical photometry, while the UV-optical 
spectra are fit
with white dwarf temperatures near 15,000K. These temperatures place the
accreting white dwarfs outside the instability zone for non-interacting DAV
white dwarfs and show that the instability strip is complex for
accreting white dwarfs. 
\end{abstract}


\section{Introduction}

Since the discovery of the first pulsating white dwarf in an accreting
close binary cataclysmic variable (GW Lib; Warner \& van Zyl 1998, van
Zyl et al. 2000, 2004), photometry
of systems with similar orbital periods and optical spectra has revealed
at least 10 more (Warner \& Woudt 2004, Woudt \& Warner 2004, 
Araujo-Betancor et al. 2005, Vanlandingham et al. 2005, 
Patterson et al. 2005a,b, Mukadam et al. 2006, 
G\"ansicke et al. 2006, Nilsson et al.
2006).
 Seven of these pulsating accreting white dwarfs were found by followup work on
cataclysmic variables (CVs) that were discovered in the Sloan Digital Sky Survey
(SDSS; York et al. 2000) by Szkody et al. (2002b,2003,2004,2005,2006). 
The optical clue for candidate pulsating white
dwarfs is the
presence of broad absorption lines (originating from the white dwarf)
flanking the Balmer emission lines (originating from the disk). In order
to be able to view the white dwarf, the accretion disk contribution to
the optical light has to be small, a situation that occurs at very low
mass transfer rates, and hence short
orbital periods (near 80 min). Since the SDSS provides spectra that
are about 2 magnitudes fainter than previous surveys, up to 25\% of
SDSS discovered CVs show the white dwarf presence (Szkody et al. 2005).
As the analysis of pulsation frequencies can lead to mass determination
and other aspects of the internal structure of the white dwarf 
via the technique of asteroseismology, the compilation of a significant number of 
pulsating CVs
provides a valuable database for probing the effects of mass transfer and
accretion on the long term evolution of white dwarfs as well as the effect
of external heat input, He enriched envelopes and fast rotation on the
location of the instability strip. 

Non-interacting hydrogen atmosphere (DA) pulsating white dwarfs (DAVs or ZZ 
Ceti stars)
 show typical non-radial g-mode pulsations with periods of 50-1400s
and have temperatures in the narrow range of 10800-12,300K and log g$\sim$8
(Bergeron et al. 1995,2004, Koester \& Allard 2000, Mukadam et al. 2004, Gianninas et al. 2005). While the pulsation frequencies
and amplitudes of CV pulsators appear to be similar to those of ZZ Ceti systems (van Zyl et 
al. 2004, Warner \& Woudt 2004, Woudt \& Warner 2004, Mukadam
et al. 2006), the temperatures of the underlying white dwarfs may be
quite different. UV observation of GW Lib (Szkody et al. 2002a) with the
Space Telescope Imaging Spectrograph (STIS) revealed a white dwarf with
a temperature of 14,700K (or a best fit with 63\% of the white dwarf
having a temperature of 13,300K and the remainder at 17,100K),
temperatures far outside the empirical instability strip for non-interacting
ZZ Ceti stars. Townsley, Arras \& Bildsten (2004) could match the pulsation periods
and high temperature if they assumed ${\ell }$=1 g-modes and a 1M$_{\odot}$
white dwarf with an accreted H-rich layer of 3$\times$10$^{-5}$M$_{\odot}$.
On the other hand, Araujo-Betancor et al. (2005) used a snapshot STIS
spectrum to show that the CV pulsator HS2331+3905 could be fit with a
10,500K, 0.6M$_{\odot}$ white dwarf, within the range of normal ZZ Ceti stars. 
Of course, pulsating CV white dwarfs are rotating very rapidly compared to
non-interacting ZZ Ceti stars and the thermal profiles are quite different
because the surface layer is heated from above by compression and 
irradiation due
to accretion. Although the accretion rates for the CV pulsating white
dwarfs are small ($\sim10^{-13}\,{\mathrm{M}}_{\odot}$ yr$^{-1}$), the thermal
profiles may still be altered and 
the effective temperature
might not be the best parameter to locate the instability strip. 
This
does not imply that the temperature in the driving regions is not important
 in locating the instability strip, but only that the effective 
temperature may not
be reflecting the temperature in the driving regions due to accretion 
heating
in the surface layers.
Arras et al.
(2006) show that the temperature in the driving zone, log g and the He
abundance are all important for accreting white dwarfs.

In order to explore the physical parameter space occupied by
 accreting pulsators (Arras et al. 2006),
 and to attempt mode identification
through comparison of UV to optical amplitudes (Robinson et al. 1995), 
we proposed further study of
three additional known pulsators with STIS. 
 After the proposal was
accepted, STIS was taken out of service but we were able to switch our program 
to the Solar Blind Channel (SBC) which has the same wavelength coverage,
albeit much lower resolution and no TIME-TAG mode of photon counting. The
three objects are SDSSJ013132.39-090122.3 (Szkody et al. 2003; Warner \&
Woudt 2004), SDSS J161033.64-010223.3 (Szkody et al. 2002; Woudt \& Warner
2004) and SDSS J220553.98+115553.7 (Szkody et al. 2003; Warner \& Woudt 2004).
Throughout the rest of this paper, we will refer to these objects as
SDSSJ0131, SDSSJ1610 and SDSSJ2205. A summary of the orbital periods,
visual brightness and optical pulsation periods is provided for these
3 systems in Table 1. The SDSS spectra are shown in Figure 1.

\section{Observations and Reductions}
\subsection{HST Ultraviolet Observations}

The HST UV data were acquired using the prism
PR110L and the Solar Blind Channel on the Advanced Camera for
Surveys (ACS) to provide
UV spectra from approximately 1245 to 1800\AA. Due to the prism,
the dispersion varied from about 1.5\AA\ pxl$^{-1}$ at the far UV end to about 
25\AA\ pxl$^{-1}$ 
at the longest wavelengths.
Each system was observed for 5 HST orbits using 61s integrations in ACCUM mode. The
dead time between observations was 40s so the time resolution of the spectra
are 101s. The first orbit had 26 integrations (due to the initial overhead
of setting up on a target) and the remaining 4 orbits contained 29 integrations.
Thus, the total time on each source consisted of 142 integrations of 61s
each or 8662s.

The UV data were analyzed using tools available under STSDAS and
pyraf (aXe14). The primary target was extracted using a variety of extraction
widths to optimize the S/N of the resulting spectra and light curves. For
the spectra, a wide extraction of 17 pixels was used and the resulting
142 spectra were co-added to produce a final spectrum. For the light curves,
a smaller extraction width produced the optimum (largest) pulse signal.  This
optimum extraction width was determined to be 9, 12 and 4 pixels for SDSSJ0131, 1610 and
2205 respectively. The fluxes were then added across 
wavelength for each
individual spectrum to produce photometric points
throughout the 5 orbits for a light curve that could be analyzed
for periodicity using Discrete Fourier transforms. For the 2 brightest
systems (SDSSJ0131 and 1610), the spectra at the peaks
and troughs of the resulting light curves were then individually combined 
into
peak and trough summed spectra, as any difference in temperature between
these spectra would help constrain the change in temperature during a
pulsation cycle of the dominant mode.

\subsection{Optical Observations}

In order to prevent excessive UV light from entering the ACS detectors,
STScI required continual ground observations prior to and during the
HST observations to insure that the objects were not going into a dwarf
nova outburst. These monitoring observations were kindly provided by a large
group of worldwide amateurs and professionals (from AAVSO and
ROTSE and scientists at observatories for other programs). All 3 systems
remained at quiesence throughout the planning and execution of the
observations. 

Coordinated simultaneous optical observations were achieved for
SDSSJ1610 in order to obtain the amplitude and period of the optical
pulsations. The 3.5m telescope at Apache Point Observatory (APO) 
was used with a B filter on the CCD camera SPIcam 
along with the Nordic Optical Telescope (NOT) 2.56m telescope and B filter on
the Faint Object Spectrograph and Camera (ALFOSC) on the night of 
June 30/July 01. APO used an integration time of 20s while NOT used 14.1s. 
Additional data were obtained on the 2m Himalayan Chandra Telescope (HCT)
on the nights preceding and following the HST observation. These observations
were accomplished with no filter and 30s integeration times.

To check the stability of periods for SDSSJ0131, additional APO data
were obtained on 2005 Dec 1 and 5 and also on 2006 Jan 30, using the imaging CCD on the
Dual Imaging Spectrograph (DIS) with no filters.
DIS utilizes a dichroic that splits the light at 5550\AA\ so that the blue
and red portions of the beam are then incident on two distinct CCD cameras. The
light curves obtained using the blue CCD camera are used in this paper, sensitive
in the wavelength range 3500-5550\AA. Windowing was used to read a small portion of the
CCD in order to reduce the read out time and obtain a suitable time resolution. 
A standard IRAF\footnote{{IRAF (Image
 Reduction and Analysis
Facility) is distributed by the National Optical Astronomy Observatories, which
are operated by AURA,
Inc., under cooperative agreement with the National Science Foundation.}}
reduction to extract sky-subtracted light curves from the CCD frames using weighted
circular aperture photometry (O'Donoghue et al. 2000) was employed.
The optical light curve of the target star was divided by a sum of one or more brighter comparison
stars and converted to the same fractional
amplitude scale and the times converted to Barycentric
Coordinated Time (Standish 1998). A Discrete Fourier
Transform (DFT) was then computed for all the optical light curves up to the Nyquist frequency.

Additional optical data for SDSSJ0131 over a longer timescale were obtained
from the 1.9m Radcliffe telescope at the Sutherland site of the South African
Astronomical Observatory (SAAO), using the University of Cape Town
CCD Photometer (UCTCCD; O'Donohue 1995) with no filter. Data were acquired during
several nights in 2003 July, August and September; 2004 September and November
and 2006 August. Fourier transforms were computed for each
night of data, as well as combined FTs for each month. 

A summary of the HST and ground-based observations is given in Table 2.

\section{Light Curves and Pulsations}

The summed photometry from each HST spectrum was analyzed in a similar manner
as the DIS data.
Figure 2 shows a light curve for a single HST orbit for each
system; the pulsation is apparent in each case. The combined data from all
5 orbits were then
analyzed with DFT routines to obtain the quantitative periods and amplitudes.
Figure 3 shows the amplitude plots and Table 3 summarizes the resulting periods
and amplitudes. The APO, NOT, and HCT data were analyzed in the same way and
the results shown in Table 3 and Figure 4.

The optimum dataset exists for SDSSJ1610 since we can compare periods and
amplitudes from simultaneous UV and 
optical data observations.
Figure 2 shows that the UV pulse is most clearly resolved for this system.
Table 3 shows that the observed UV and optical 
periods are identical, while the UV amplitudes are
increased by 2-6 times over those of the optical. Our optical results are
consistent in period and amplitude with the 2 independent periods
that were evident in the  
past data of Woudt \& Warner (2004) as listed in Table 1. The observed
304s period is a harmonic of the 608s period.

For pulsating white dwarfs, the identification of the excited 
non-radial g-modes
can provide the total mass of the star as well as the
mass of the H layer (Kawaler \& Hansen 1989).
Each of the eigenmodes that can be excited in the star is described by a set of indices:
 $k$ is the radial quantum number that gives the number of nodes between the
surface and the center of the star,
 $\ell $ is the azimuthal quantum number that gives the number of nodes on the surface,
 and $m$ is the number of nodes along the meridian; it is used to describe the
 frequency if the spherical symmetry is lost due to rotation or a magnetic field.
 We use spherical harmonics (${{\mathrm{Y}}^{\ell }}_m$) to describe these eigenmodes.
 This identification
of the quantum numbers $k$, $\ell $, 
and $m$ for each observed period
is not easy, as it involves matching observed periods with predictions from
models (with many free parameters). Robinson et al. (1995) 
pioneered a method of determining the $\ell $ value
based on the change in amplitude as a function of wavelength
due to limb darkening and the modified geometric cancellation (Robinson,
Kepler \& Nather 1982).
 Limb darkening effectively reduces the viewing area of the stellar disk, 
 and this reduction in area depends on wavelength.
 At UV wavelengths, the increased limb darkening
 decreases the contribution of zones near
the limb, and modes of $\ell $=3 are canceled less effectively in the UV 
compared to $\ell $=1 or 2.
However, this does not hold true for $\ell $=4 modes, which do not show a significant
change in amplitude as a function of wavelength.

 Robinson et al. (1995), Nitta et al. (2000) and Kepler et al. (2000)
 calculated
amplitudes in the UV for different values of $\ell $.  To attempt to
constrain the $\ell $ value for the 608s period observed in SDSSJ1610, we 
determined the amplitudes for this period
for four different wavelength regions: 1245-1360\AA, 
1360-1500\AA, 1500-1640\AA, and 1640-1800\AA. The effective wavelengths
of these regions were computed using the best-fit white dwarf model
(section 4). We also computed the amplitudes for 20 individual wavelength bins 
but the
resulting error bars were too large to be useful. Figure 5 shows the
observed pulse amplitude ratios for our four wavelength regions
 along with the theoretical
amplitude ratio for single DA white dwarf stars using Koester's atmosphere
models (Finley, Koester \& Basri 1997). The top panel shows the model
calculations for $\ell $=1 to 4 modes for log g=8, T$_{eff}$=14,500K (top) and 12,500K
(bottom). 
 While neither plot can
fit all the wavelengths, the $\ell $=1 mode with T$_{eff}$=12,500K is a
good fit to all but the shortest wavelength and it appears that high
order modes ($\ell $=3,4) can be ruled out.
However, since
the temperatures for accreting white dwarfs are quite different than for DAVs 
(see below), the simple non-interacting DA
model may not be applicable. In addition, since the disk may
cast a shadow on the equatorial regions of the white dwarf, the limb
darkening and geometric cancellation may be different than that for single 
white 
dwarfs. 

It is also interesting that the peaks in the FT of SDSSJ1610 show
no splitting (Figure 4), even though the resolution is $\sim$10$^{-5}$ Hz 
(vsini$\sim$0.5 km s$^{-1}$),
whereas fitting of line widths in HST spectra
have shown velocities about 50-100 km s$^{-1}$ (Sion et al. 1994) for accreting
white dwarfs. However, there may be some process that is suppressing some of the modes.
Long
term monitoring of GW Lib (van Zyl et al. 2004) showed  clusters of
periods separated only by about 1 $\mu$Hz whose origin is not clear. It
will take a much longer combined run on SDSSJ1610 to tell if this is also
the case in this system.

For the other two objects, where simultaneous optical light curves could not be
obtained, we can only compare the observed UV periods and amplitudes with
available optical values. For SDSSJ2205, the single observed
UV period of 576s matches one of the  previously observed optical periods
noted by Warner
\& Woudt (2004), and the UV amplitude is about 6 times the optical value. The
strongest period (330s) observed in the optical by Warner \& Woudt is
not visible in the UV.

The brightest system (SDSSJ0131) shows no UV pulsation at the
highest amplitude period (595s) previously evident in the optical 
data published in  Warner \& Woudt
(2004). Their other 2 optical periods (260s and 335s) are also not
apparent in the SBC data. 
Instead, a shorter period of 213 s near the Nyquist frequencey 
is evident in the HST data
with an amplitude near 80 mma.
The 213s period
is obvious, although just barely resolved, in the SBC light curve (Figure 2).
This is the shortest period yet seen in an accreting pulsator. 
Our extended optical coverage on this object with APO in the months following the HST 
observation and with SAAO in the years preceding and following the HST
(Table 2) provides a handle on the stability of the observed periods.
Figure 6 shows the DFT of the 2006 January data while Figure 7 shows all
the monthly combined runs from the SAAO (Table 4 lists the observed
frequencies and periods from the SAAO runs). It is clear that the pulsation
spectra are highly variable, with longer periods most prominent in the
runs of 2003 (Warner \& Woudt 2004 use data from 2003 September), while
the 2004 runs show only the short period seen in the 2005 HST data.  Then, in
2006 January (Figure 6) both the long period (near 600s) and the short (213s)
period are evident, and in 2006 August (Figure 7), these are both still
present, along with an even longer period (1130s) that is similar to one
seen in 2003 Aug.
While
changes in pulsation spectra are common in cool non-interacting ZZ Ceti
stars on timescales of months and even days,
it is not obvious what is to be
expected with these hotter pulsators. Long term monitoring of GW Lib
(van Zyl et al. 2004) does show that its pulsation amplitudes are unstable
as well. The 2 periods near 650 and 370s remain visible
in all runs (including the HST data), while other periods appear and become
invisible in other runs. 
In SDSSJ0131, it appears that
all periods show large amplitude variations. Comparing the amplitude of the 213s period in
the UV with that seen in the optical gives a ratio of about 6, similar to
that of SDSSJ1610 and 2205.

\section{Spectral Fits}

The UV spectra of low mass transfer rate CVs are a combination of
the underlying white dwarf, the accretion disk and a source of emission
lines (likely an accretion disk chromosphere). To model the observed
SBC spectra, we employed a procedure similar to our analysis of STIS
data on similar systems (e.g. G\"ansicke et al. 2005). This involved using
Hubeny white dwarf LTE models TLUSTY195 and SYNSPEC46 (Hubeny \& Lanz 1995) 
in a grid
of temperatures, gravities and abundances to create a best fit to the
spectrum. Due to the low resolution of the SBC and the past results which were
all consistent with log g=8, we fixed the log g at this value and used 0.01
solar abundance to fit the best temperature white dwarf. The constraints on
the fit include that the white
dwarf must provide the best match to the UV continuum shape, especially the broad
Ly$\alpha$ line, and its optical flux must fall below the observed optical 
fluxes. Figure 8
shows the model results for a range of white dwarf temperatures (10,000K to
20,000K) compared to the observed HST and SDSS spectra of the brightest
system SDSSJ0131. It is immediately clear that the temperature typical
of non-interacting
ZZ Cet pulsators ($\sim$12000K) is too cool to fit the UV spectrum. A temperature
of 14,500K is derived from the best fit. This temperature white dwarf extrapolated to
optical wavelengths produces a $\it g$ magnitude of 19.2, well below the 
observed SDSS $\it g$
mag of 18.3. In addition to the white dwarf, there is some continuum source
(the disk) that produces about 30\% of the light across the spectrum. This
is especially evident in the core of Ly$\alpha$ which should be zero for
a high gravity white dwarf.  
As there is no good model of a non-steady
state disk at these low mass transfer rates, we added this continuum
component as a simple black body (power laws were tried with similar results),
used simple Gaussians to fill in the emission lines and redid the fits with
the two components. The resulting best fit for SDSSJ0131 is shown in Figure
9 and the results summarized in Table 5. A similar procedure produced the
best fits for SDSSJ1610 and SDSSJ2205 (Figures 10 and 11 and Table 5).

The spectra that produced the largest amplitude pulse were also added and fit 
with
this procedure. These spectra had higher S/N but less overall flux. Thus,
the shape of the spectrum may be better defined but the total fluxes are
not correct for distances, etc. We also list these temperature fits as
T$_{optext}$ in Table 5
as a measure of the error of our temperature fits.

We also attempted to fit the difference between the peaks of the 
pulsations and the troughs for the two brightest sources SDSSJ0131 and
SDSSJ1610. The spectra of the peak points in the light curves were summed
to produce one total peak spectrum
and then the troughs were summed to produce a total trough spectrum. 
The resulting fits to the peak and trough spectra showed
no change in the white dwarf temperatures in either case,
which is not surprising 
given that the uncertainties of our fits are on the order of 
1000K. 

The temperatures of the white dwarf in all 3 systems are similar,
close to 15,000K and much hotter than the temperatures of the DAV
pulsators. These temperatures are also very similar to that derived for
GW Lib (Szkody et al. 2002). Because we had used a two-temperature white dwarf
fit for GW Lib, rather than a white dwarf + BB, we redid the
fits to the STIS data of GW Lib using the approach here. The result
produces a best fit with T$_{wd}$=15,400K and T$_{BB}$=12,000K but
this fit is not as good as our past fit with 2 white dwarf components of
13,300K and 17,100 covering 63\% and 37\% respectively 
of the white dwarf surface. Further confirmation that the two temperature white dwarf
is correct for GW Lib comes from inspection of its STIS spectrum. This  
shows that the core
of Ly$\alpha$ does go down to zero flux, indicating that the white dwarf, rather
than a second component such as the disk, contributes all the UV flux. A
comparison of the single temperature white dwarf fit (14,700K) with that of
the WD+BB also shows that adding in a second component only changes the
derived temperature of the white dwarf by less than 1000K. 

Thus, we now have secure results that four of the accreting ZZ Ceti stars
in CVs are hotter than single pulsators. On the other hand, HS2331+3905
is also secure in having a white dwarf temperature of 10,500K (see
Fig. 13 in Araujo-Betancor et al. 2005). While the UV emission lines in
HS2331+3905 are much stronger than in the four objects that have hot white
dwarfs (likely partly due to the weak white dwarf continuum), and 
there are continuum variations which might imply a higher
inclination, the fact remains that the white dwarf pulsations are visible
in HS2331+3905 so the white dwarf light is prominent and it is cool.
In contrast to our three SDSS sources, all of which have UV fluxes greater
than the optical as in Figure 8, HS2331 has larger optical than UV fluxes, 
and a distinct rise in flux longward of 1600\AA\ which pins down the
white dwarf temperature to its cool value.

Since accreting white dwarfs are known to show UV absorption lines of
Si, C and other metals, it is clear they do not have pure H atmospheres 
(Sion 1999). Thus, perhaps it is not unexpected that the instability strip
is not the same as for ZZ Ceti stars. For accreting model white dwarfs
with a high He abundance ($>$0.38), Arras, Townsley \& Bildsten (2006) 
find an additional hotter instability strip at $\sim$15,000K for low mass
white dwarfs due to \ion{He}{2} ionization.
 Thus, they
infer that HS2331+3905 has a low mass white dwarf while GW Lib (and hence 
SDSSJ0131, SDSSJ1610 and SDSSJ2205) should have highly evolved donor stars and more massive
white dwarfs.

\section{Conclusions}

Our low resolution light curves throughout five HST orbits for our 
three white dwarf accreting pulsators all have increased pulsation amplitudes
in the UV compared to the optical, consistent with $\ell $=1 modes. 
The simultaneous optical/UV data
for SDSSJ1610 showed identical periods in both wavelength regions, while
SDSSJ0131 and SDSSJ2205 showed one of three periods evident in the optical
at other times. The long term monitoring of SDS0131 shows that all pulsation
periods come and go, indicating a high level of instability for this system. 

The summed UV spectra reveal hot white dwarfs in all three systems, similar
to past results on GW Lib but contrary to the lone cool accreting white dwarf
pulsator HS2331+3905. Based on just these five systems with well-determined
white dwarf temperatures from the UV, it appears that there is a wide
range in the instability strip for accreting systems, with most accreting
pulsators in a
regime that is hotter than for non-accreting ZZ Ceti stars. Whether this is due to
a difference in the white dwarf mass and/or He composition provided by an
evolved donor as proposed by
Arras, Townsley \& Bildsten 2006 can be confirmed if the white dwarf
masses and/or the
composition of the donor can be determined. As there is no evidence of
a donor star out to 9000\AA\ in the SDSS spectra and the systems are
generally faint, the donor will be difficult to characterize. Since the
absorption lines of the white dwarf are difficult to measure in the optical
with the contaminating Balmer emission, the best approach may be the use of
high resolution UV (the Cosmic Origins Spectrograph) time-resolved data
in the future to obtain the mass of the white dwarf, although the unknown inclinations will
create some uncertainty.

\acknowledgments

We are very grateful to the many observers who monitored our objects
to insure the safety of the UV detectors and allow our
observations to take place, including Ms. Bama, Dick Campbell, Dean Chandler, Katy Garmany 
(and TLRBSE teachers), Bill Goff, Bernard Heathcote, Arne
Henden, Santosh Joshi, Seung-Lee Kim, Mercedes Lopez-Morales, David Mary, 
John W. McAnally, Aaron Price, Chuck Pullen, ROTSE observers,
Patrick Schmeer, and D. E. Winget. We also kindly acknowledge Denise Dale and Ewald
Zietsman for the observations of SDSS J0131 taken at the SAAO.
This research was supported by NASA grant GO-10233.01A and through the
Hubble Fellowship grant HST-HF-01175.01-A, both awarded from the Space
Telescope Science Institute which is operated
 by the Association of Universities for Research in Astronomy, Inc., for NASA,
 under contract NAS 5-26555.
BTG was supported by a PPARC Advanced Fellowship. 
The NOT data were taken using the ALFOSC, which is owned by
the Institute de Astrofisica de Andelucia (IAA) and operated by the Nordic
Optical Telescope under agreement between IAA and the NBIfAFG of the
Astronomical Observatory of Copenhagen.

\clearpage
\begin{deluxetable}{lcclll}
\tablewidth{0pt}
\tablecaption{Source Characteristics}
\tablehead{
\colhead{SDSS J} & \colhead{{\it g}(mag)} & \colhead{P$_{orb}$(min)} &
\colhead{Opt Pulse P (sec)} & \colhead{Amp (mma)} & \colhead{Ref} }
\startdata
0131 & 18 & 98 & 260, 335, 595 & 4, 9, 17 & Warner \& Woudt 2004 \\
1610 & 19 & 81 & 345, 607\tablenotemark{a} & 7, 27 & Woudt \& Warner 2004 \\
2205 & 20 & -- & 330, 475, 575 & 10, 7, 8 & Warner \& Woudt 2004 \\
\enddata
\tablenotetext{a}{Several combinations of these 2 independent periods are
also evident.}
\end{deluxetable}
\clearpage

\begin{deluxetable}{ccllrl}
\tablewidth{0pt}
\tablecaption{Summary of Observations}
\tablehead{
\colhead{SDSSJ} & \colhead{Obs} & \colhead{Instr} & \colhead{Filter} & \colhead {Int(s)} & \colhead{Time} }
\startdata
0131 & HST & SBC &  PR110L & 61 & 2005 Jun 18 1:35:49\,-\,8:42:35 \\
0131 & APO & DIS & none & 15, 25 & 2005 Dec 01 07:54:56\,-\,08:42:22 \\
0131 & APO & DIS & none & 15, 30 & 2005 Dec 05 07:25:21\,-\,08:19:53 \\
0131 & APO & DIS & none & 15 & 2006 Jan 30 01:41:23\,-\,03:12:27\\
0131 & SAAO & UCTCCD & none & 30 & 2003 July 2 (2.18h), 3 (2.23h), 5 (2.17h) \\
0131 & SAAO & UCTCCD & none & 120, 100 & 2003 July 25 (3.34h), 28 (3.47h) \\
0131 & SAAO & UCTCCD & none & 20, 60 & 2003 Aug 20 (2.32h), 22 (1.53h) \\
0131 & SAAO & UCTCCD & none & 10-90 & 2003 Sep 17 (3.37h), 20 (1.47h), 21 (4.65h), \\
   & & & & & 22 (6.16h), 24 (1.28h), 25 (2.20h) \\
0131 & SAAO & UCTCCD & none & 60 & 2004 Sep 14 (2.88h), 15 (3.48h), 16 (5.10h) \\
  & & & & & 19 (3.13h), 20 (5.11h), 21 (0.83h) \\
0131 & SAAO & UCTCCD & none & 30-60 & 2004 Nov 3 (5.43h), 4 (1.53h), 5 (3.21h), 6 (4.57h) \\
0131 & SAAO & UCTCCD & none & 20 & 2006 Aug 25 (4.03h), 26 (4.78h), 27 (5.05h) \\
1610 & HCT & CCD & none & 30 & 2005 Jun 30 15:21:21\,-\,16:22:56 \\
1610 & HST & SBC & PR110L & 61 & 2005 Jun 30 23:13:19\,-\,2005 Jul 01 6:18:13 \\
1610 & NOT & ALFOSC & B & 14.1 & 2005 Jun 30 22:16:59\,-\,2005 Jul 01 03:29:39 \\
1610 & APO & SPIcam & B & 20 & 2005 Jul 01 03:33:33\,-\,07:58:28 \\
1610 & HCT & CCD & none & 30 & 2005 Jul 01 17:01:53\,-\,17:56:34 \\
2205 & HST & SBC & PR110L & 61 & 2005 May 23 16:05:03\,-\,23:07:33 \\
\enddata
\end{deluxetable}

\clearpage
\tiny
\begin{deluxetable}{llll}
\tablewidth{0pt}
\tablecaption{Summary of Observed Periods (HST, APO, NOT, HCT)}
\tablehead{
\colhead{SDSSJ} & \colhead{Wavelength (\AA)} & \colhead{Period (s)} & \colhead{Amp (mma)} }
\startdata
0131 & 1255-1800 & 213.72$\pm$0.10 & 78.0$\pm$8.8 \\
0131 & optical & 581.3$\pm$3.8; 211.8$\pm$1.2; 79.3$\pm$0.2 & 31.3$\pm$3.3; 13.6$\pm$3.3; 9.5$\pm$3.3 \\
1610 & 1245-1800 & 608.22$\pm$0.30; 304.10$\pm$0.29\tablenotemark{a}; 220.81$\pm$0.31 & 186.1$\pm$7.5; 48.3$\pm$7.5; 23.4$\pm$7.4 \\
1610 & optical & 608.20$\pm$0.28; 304.16$\pm$0.20\tablenotemark{a}; 220.68$\pm$0.09 & 30.6$\pm$2.2; 10.5$\pm$2.2; 12.2$\pm$2.2 \\
2205 & 1202-1800 & 576.2$\pm$1.6 & 46$\pm$11 \\
\enddata
\tablenotetext{a}{304s period is a harmonic of the 608s period.}
\end{deluxetable}

\normalsize
\clearpage
\begin{deluxetable}{lcclc}
\tablewidth{0pt}
\tablecaption{Summary of SAAO Periods for SDSSJ0131}
\tablehead{
\colhead{Run} & \colhead{Freq ($\mu$Hz)} & \colhead{Period (s)} & \colhead{Amp (mma)} & \colhead{Comment} }
\startdata
2003 July 2-5 & 1664.1$\pm$0.2 & 600.9 & 20.9$\pm$2.1 & f1\tablenotemark{a} \\
 & 2989.5$\pm$0.3 & 334.5 & 12.1$\pm$2.2 & f2 \\
 & 4629.5$\pm$0.4 & 216.0 & 8.0$\pm$2.2 & f3\tablenotemark{a} \\

2003 July 25-28 & 1690.0$\pm$0.2 & 591.7 & 20.6$\pm$3.0 & f1 \\
 & 1603.1$\pm$0.4 & 628.8 & 10.4$\pm$3.0 &  \\
 & 2970.9$\pm$0.4 & 336.6 & 9.2$\pm$3.0 & f2 \\
2003 Aug 20-22 & 861.7$\pm$0.4 & 1160.5 & 11.5$\pm$2.3 & f5 \\ 
 & 1680.8$\pm$0.2 & 595.0 & 24.6$\pm$2.4 & f1 \\
 & 2946.2$\pm$0.4 & 339.4 & 12.2$\pm$2.3 & f2 \\
 & 3777.8$\pm$0.5 & 264.7 & 9.9$\pm$2.4 & f4 \\
2003 Sep 17-25 & 1687.5$\pm$0.1 & 592.6 & 13.1$\pm$1.1 & f1 \\
 & 2965.5$\pm$0.1 & 337.2 & 8.9$\pm$1.1 & f2 \\
 & 3786.2$\pm$0.2 & 264.1 & 4.1$\pm$1.1 & f4 \\
  & 7393.0$\pm$0.2 & 135.3 & 3.8$\pm$1.1 &  \\
2004 Sep 14-21 & 4671.7$\pm$0.1 & 214.1 & 10.8$\pm$1.8 & f3 \\
2004 Nov 3-6 & 4685.3$\pm$0.1 & 213.4 & 12.7$\pm$1.3 & f3 \\
2006 Aug 25-27 & 885.2$\pm$0.2 & 1129.6 & 16.2$\pm$1.1 & f5? \\
 & 1758.9$\pm$0.2 & 568.5 & 14.7$\pm$1.1 & f1? \\
 & 4669.2$\pm0.3$ & 214.2 & 10.8$\pm$1.1 & f3 \\ 
\enddata
\tablenotetext{a}{f3 may be f1+f2?; f1 may be 2f5?}
\end{deluxetable}
\clearpage
\begin{deluxetable}{lccc}
\tablewidth{0pt}
\tablecaption{White Dwarf Fits}
\tablehead{
\colhead{Parameter} & \colhead{SDSSJ0131} & \colhead{SDSSJ1610} & \colhead{SDSSJ2205} }
\startdata
T$_{wd}$ (K) & 14500 & 14500 & 15000 \\
T$_{BB}$ (K) & 15000 & 12000 & 11000 \\
distance (pc) & 360 & 480 & 810 \\
g$_{wd}$ (model) & 19.3 & 19.9 & 21.0 \\
g$_{wd}$ (sdss) & 18.3 & 19.1 & 20.1 \\
T$_{wd}$ (optext) & 14000 & 14000 & 14000 \\
T$_{BB}$ (optext) & 16000 & 16000 & 14000 \\
\enddata
\end{deluxetable} 

\clearpage

\begin{figure} [p]
\figurenum {1}
\plotone{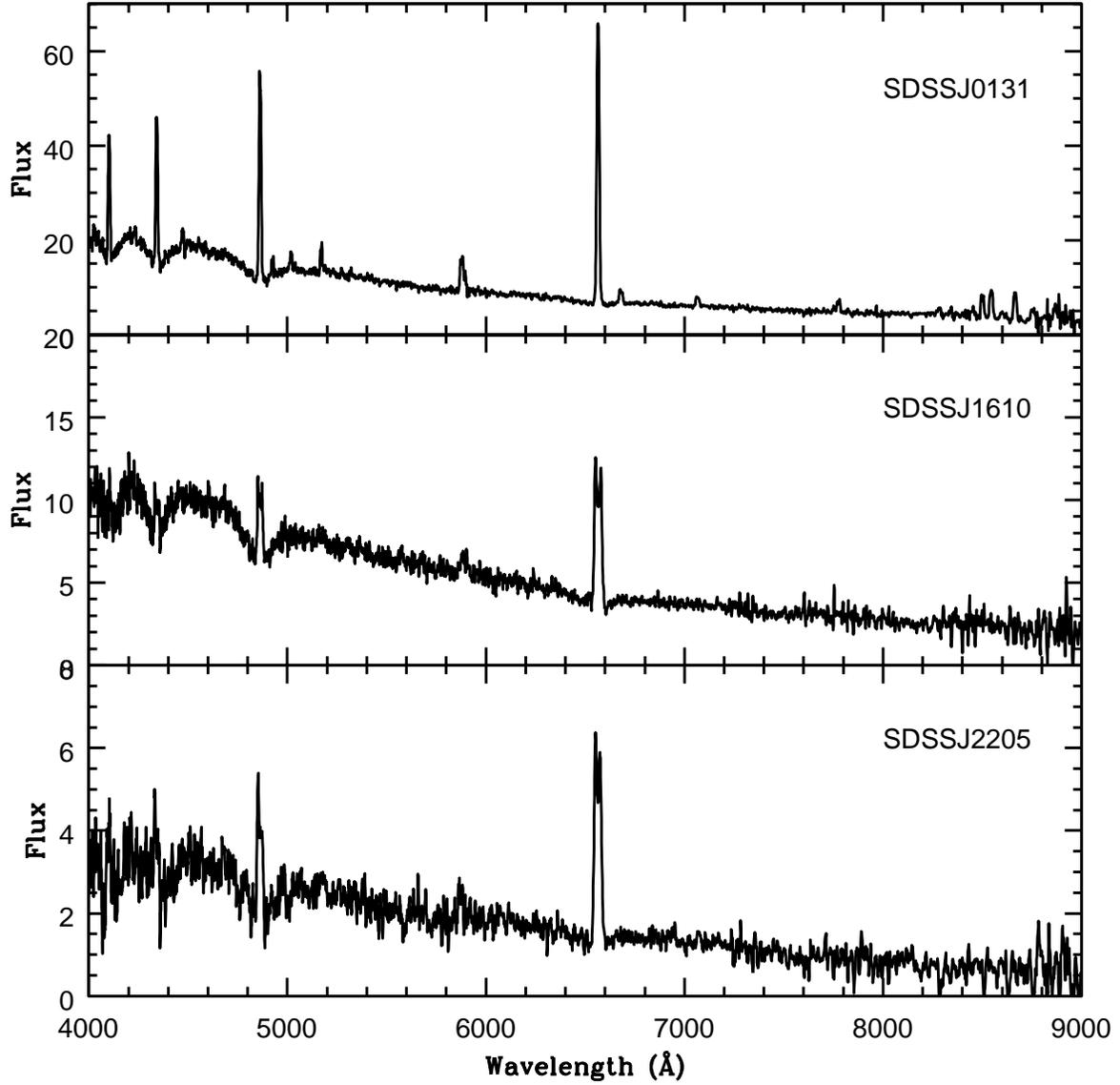}
\caption{SDSS spectra of our 3 sources, showing the absorption troughs
surrounding the Balmer emission lines. Units of flux are 10$^{-17}$ ergs
cm$^{-2}$ s$^{-1}$ \AA$^{-1}$.}
\end{figure}

\begin{figure} [p]
\figurenum {2}
\plotone{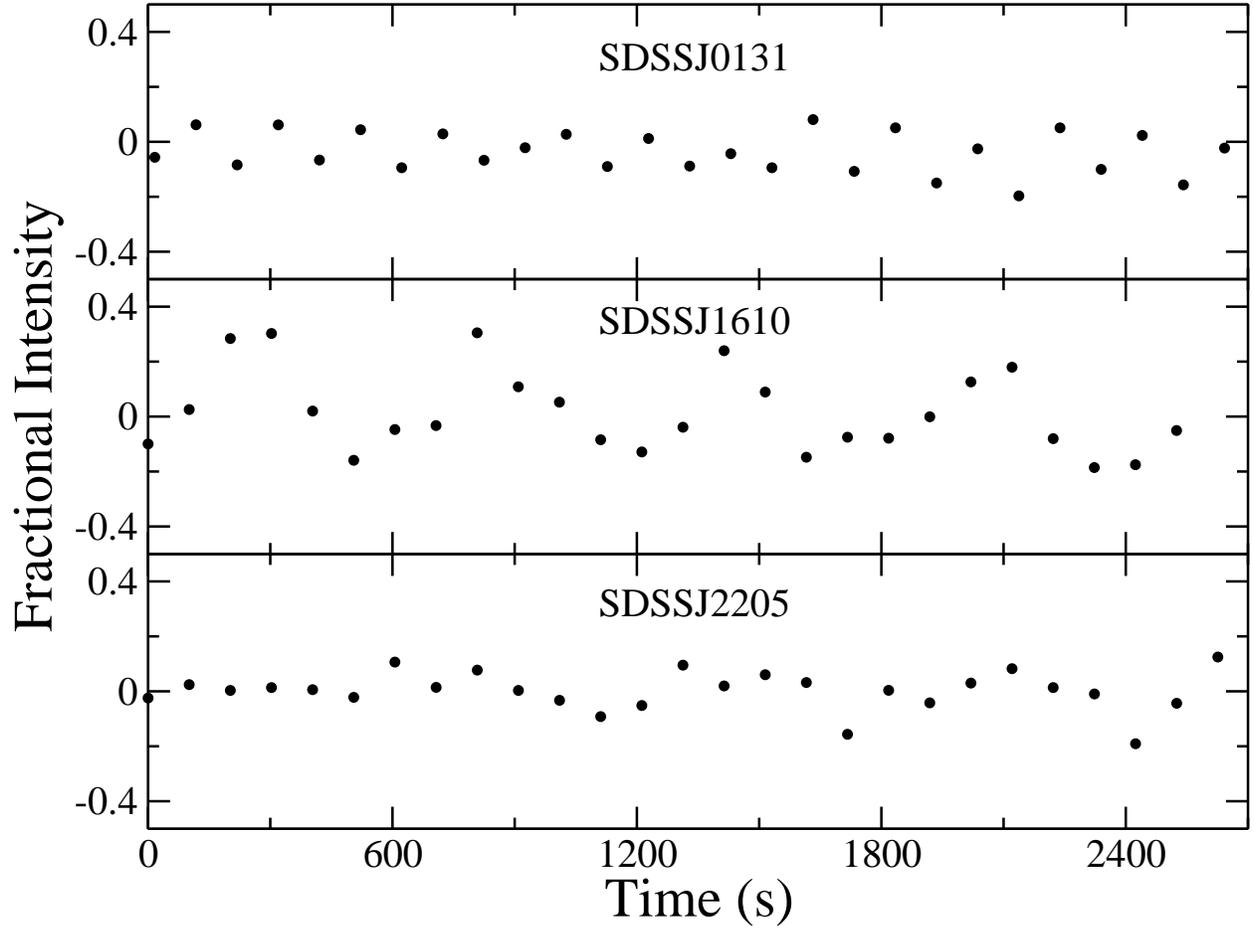}
\caption{Single orbit SBC light curves from the extracted spectra for each
of our 3 systems. Each point is one 61s integration.}
\end{figure}

\begin{figure} [p]
\figurenum {3} 
\plotone{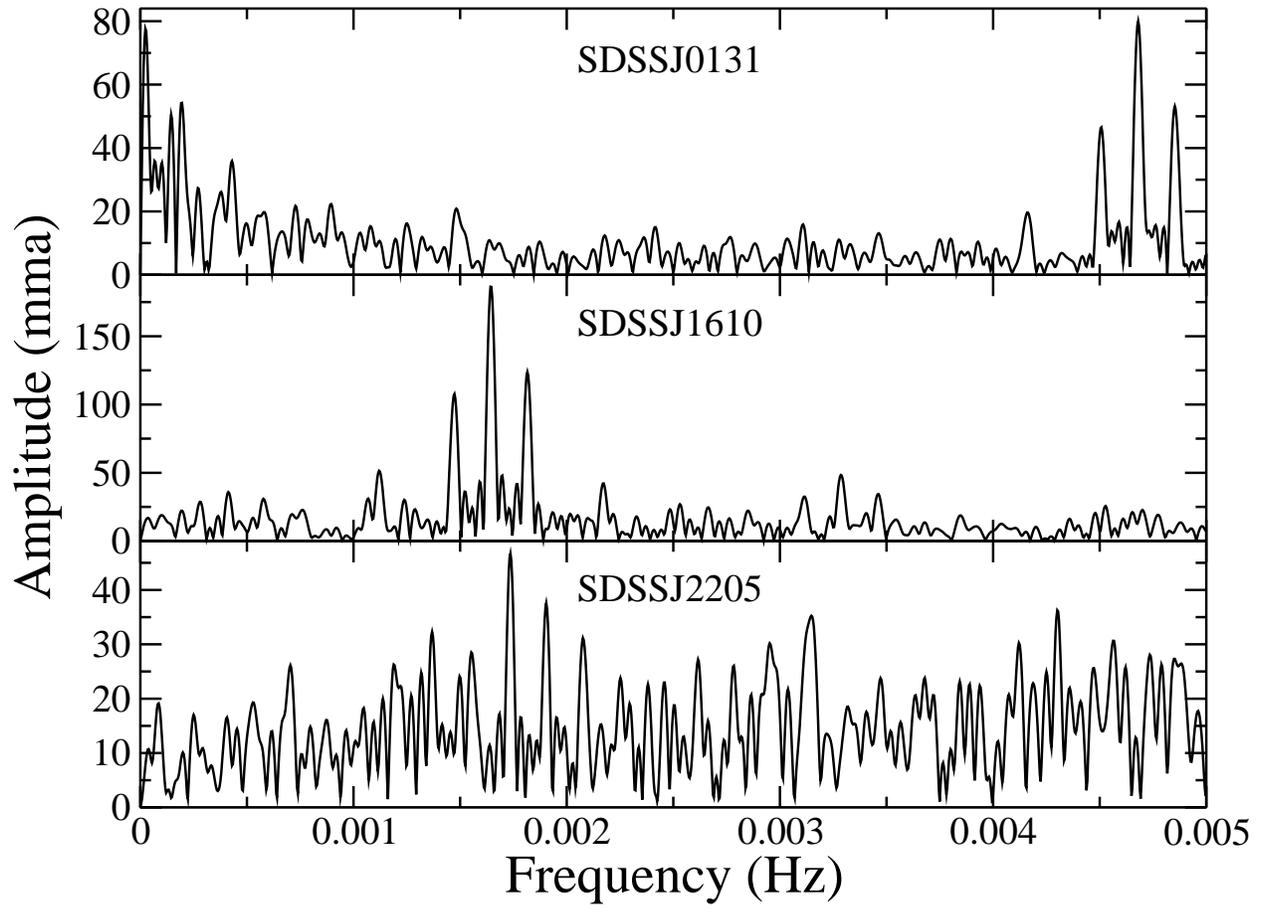}
\caption{DFTs of the combined light curves for the 5 HST orbits of each system.}
\end{figure}

\begin{figure} [p]
\figurenum {4}
\plotone{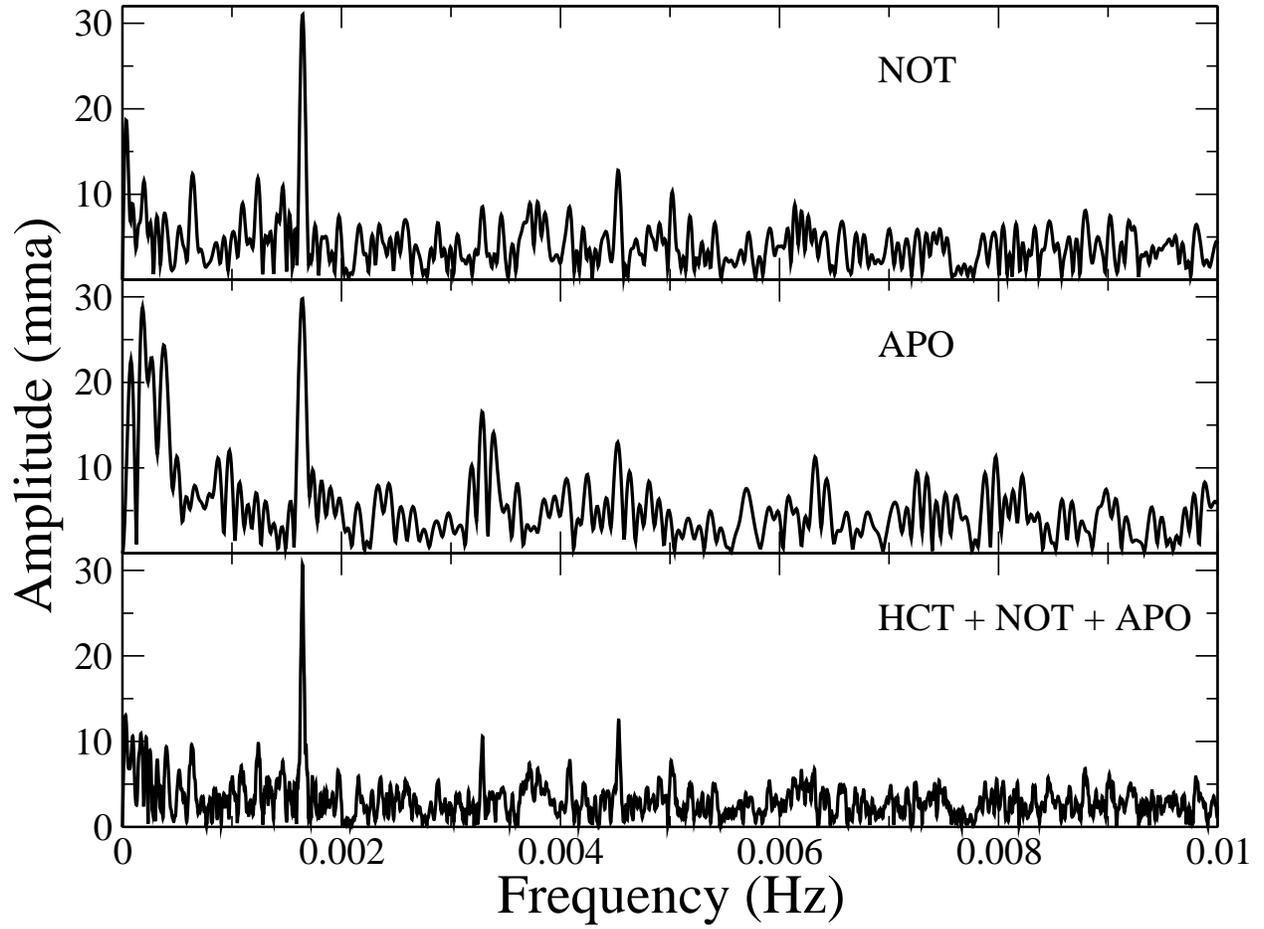}
\caption{DFTs of the APO and NOT data obtained simultaneously with the HST
observation of SDSSJ1610 and the combined DFT of the APO, NOT and HCT data.}
\end{figure}

\begin{figure} [p]
\figurenum {5}
\epsscale{0.9}
\plotone{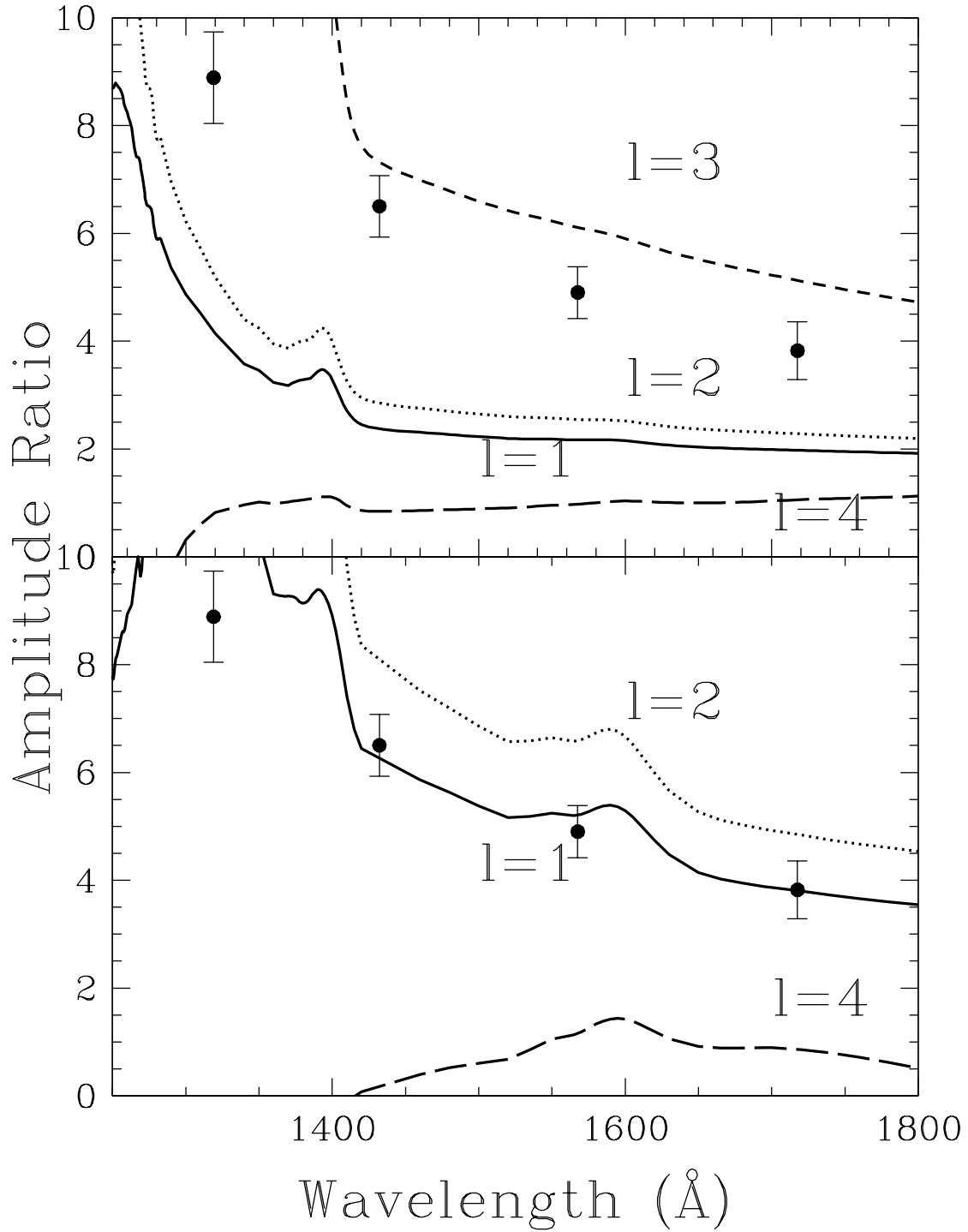}
\caption{Amplitude ratio expected for different modes of the 608s pulsation
for DA white dwarfs of log g=8 and 2 different T$_{eff}$: top, 14,500K and 
bottom, 12500K. Observed ratios at 4 different UV wavelengths sampled for
SDSSJ1610 are superposed.}
\end{figure}

\begin{figure} [p]
\figurenum {6}
\plotone{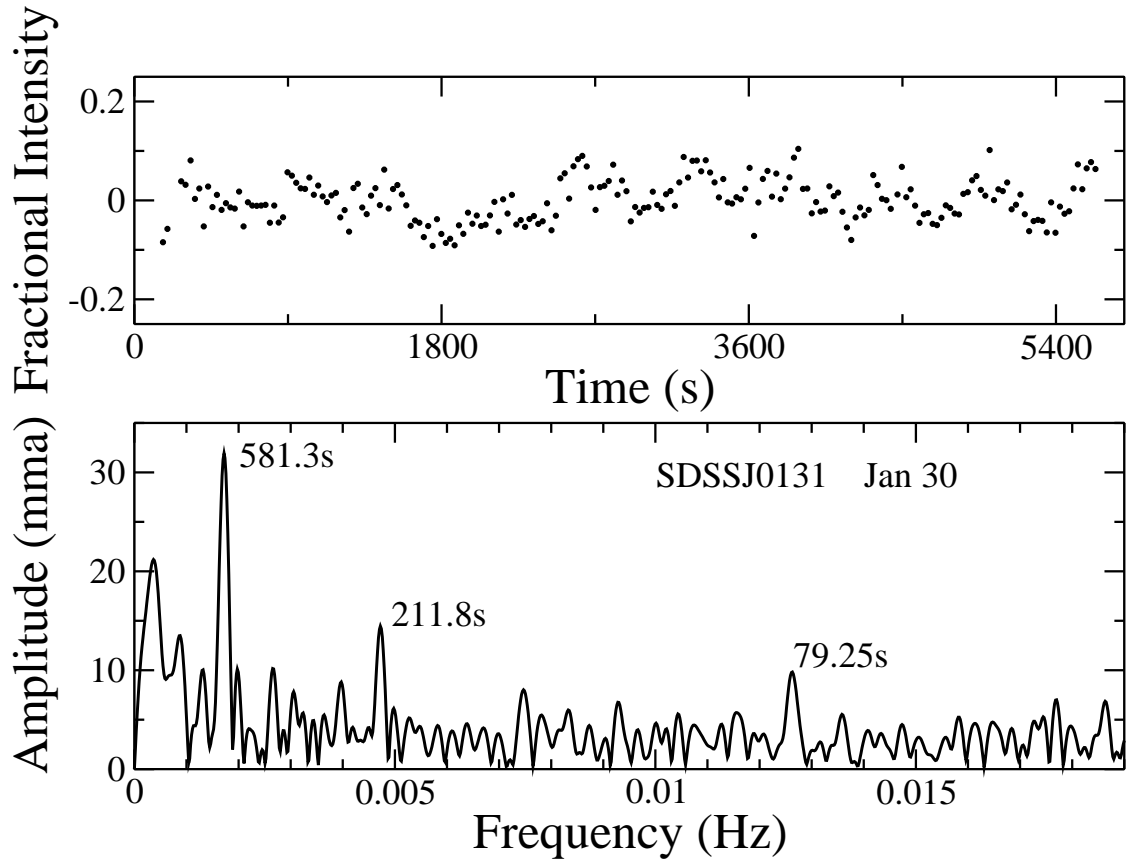}
\caption{Light curve and DFT for 2006 Jan 30 APO data on SDSSJ0131.}
\end{figure}

\begin{figure} [p]
\figurenum {7}
\plotone{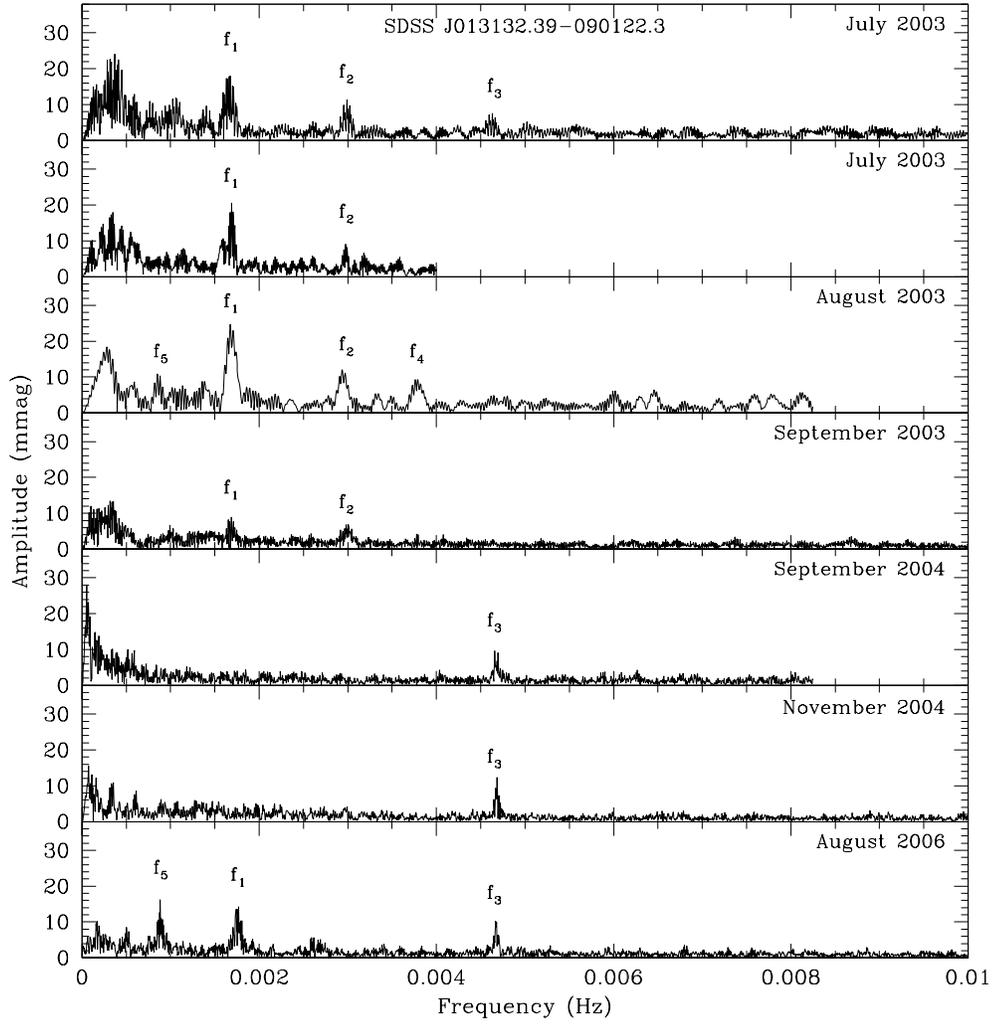}
\caption{FTs for all SAAO data on SDSSJ0131.}
\end{figure}

\begin{figure} [p]
\figurenum {8}
\vspace*{-3mm}
\plotone{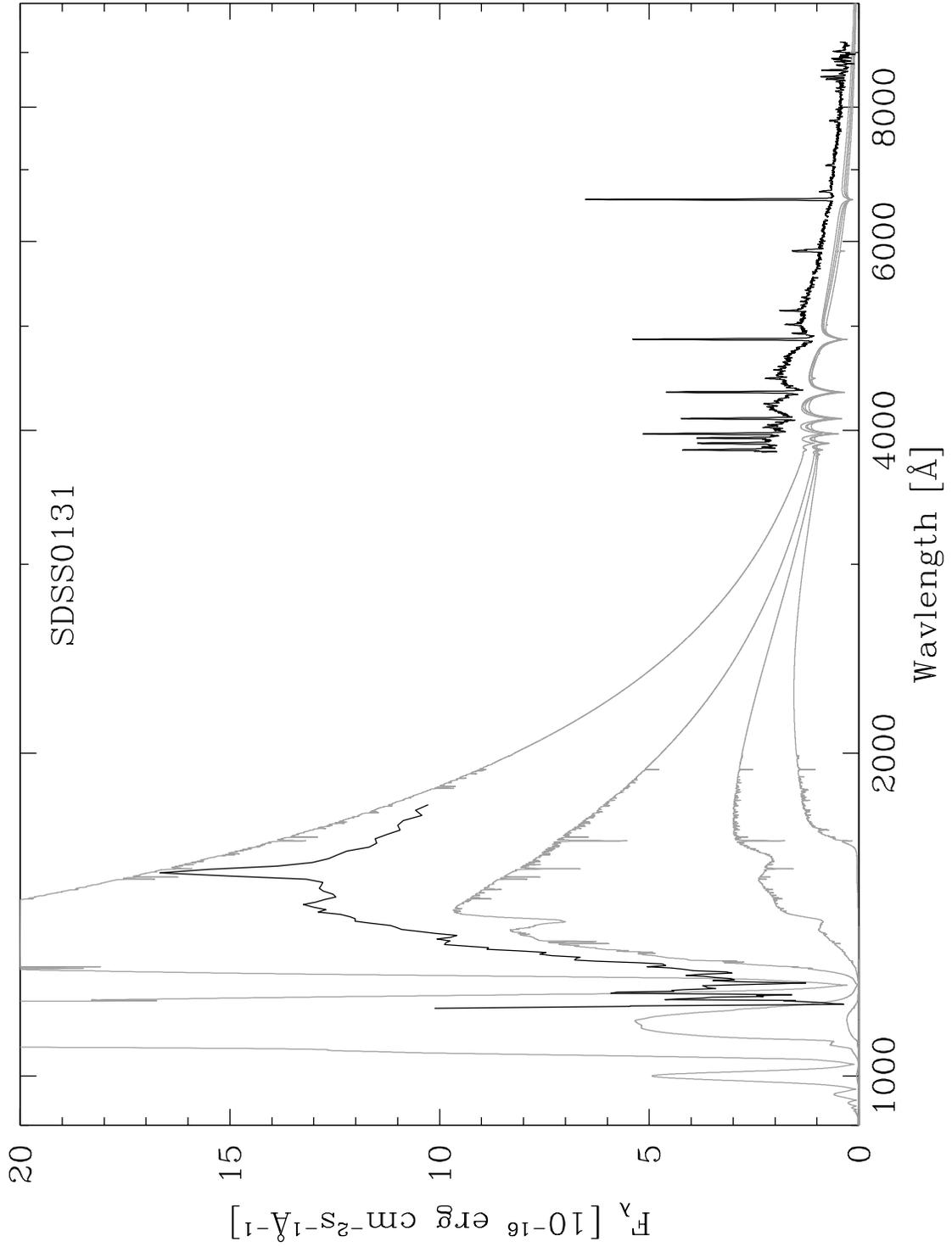}
\caption{SBC and SDSS spectra of SDSSJ0131 (dark lines) compared with model
white dwarfs from 10,000K, 12,000K, 14,000K and 20,000K (light grey lines
from bottom to top). The best fit was a 14,000K white dwarf.}
\end{figure}

\begin{figure} [p]
\figurenum {9}
\plotone{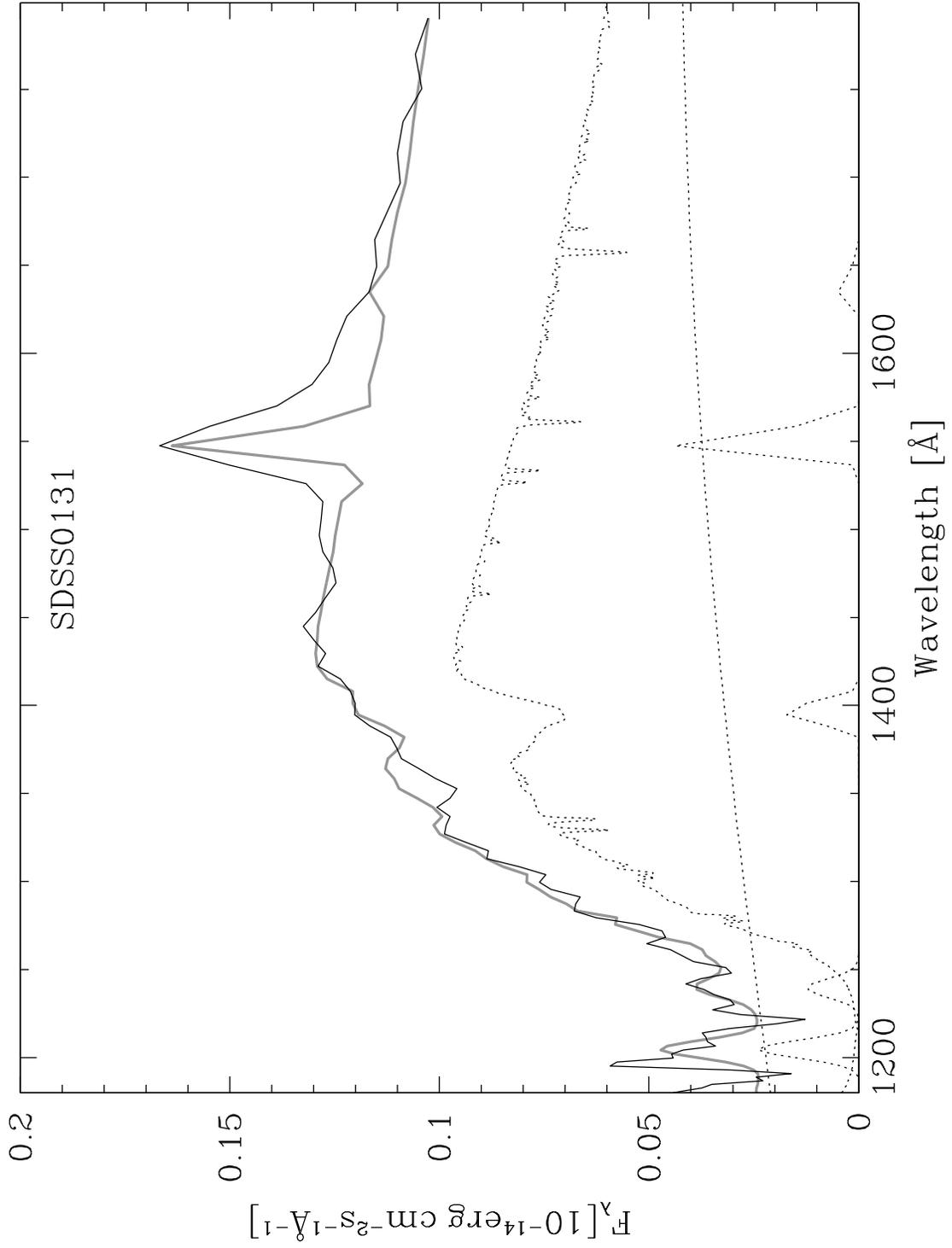}
\caption{Best fit of SBC spectrum of SDSSJ0131 with a 14,400K white dwarf at 350 pc
and 15,000K black body component.}
\end{figure}

\begin{figure} [p]
\figurenum {10}
\plotone{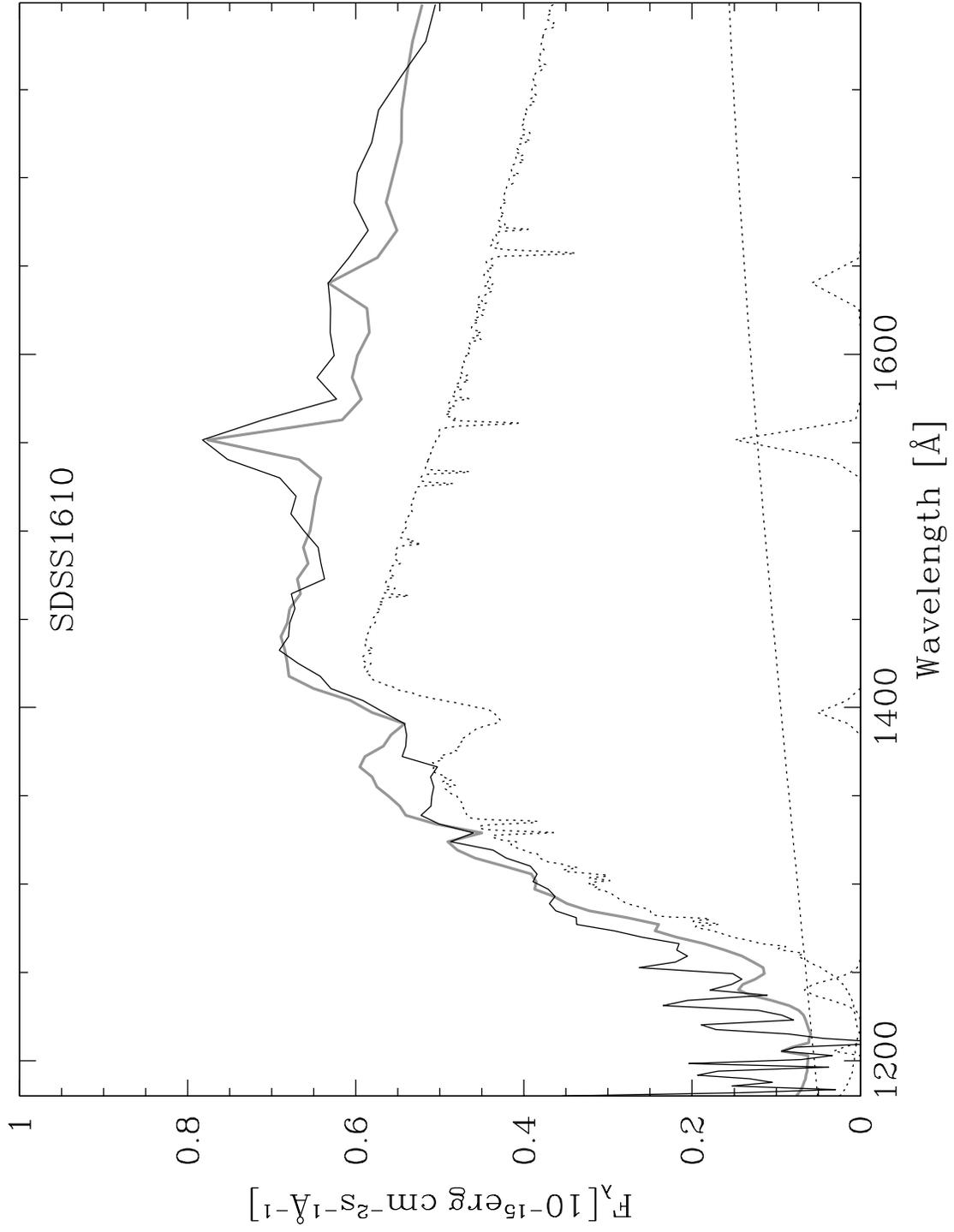}
\caption{Best fit of SDSSJ1610 with a 14,500K white dwarf at a distance of 480 pc
and a 12000K black body.}
\end{figure}

\begin{figure} [p]
\figurenum {11}
\plotone{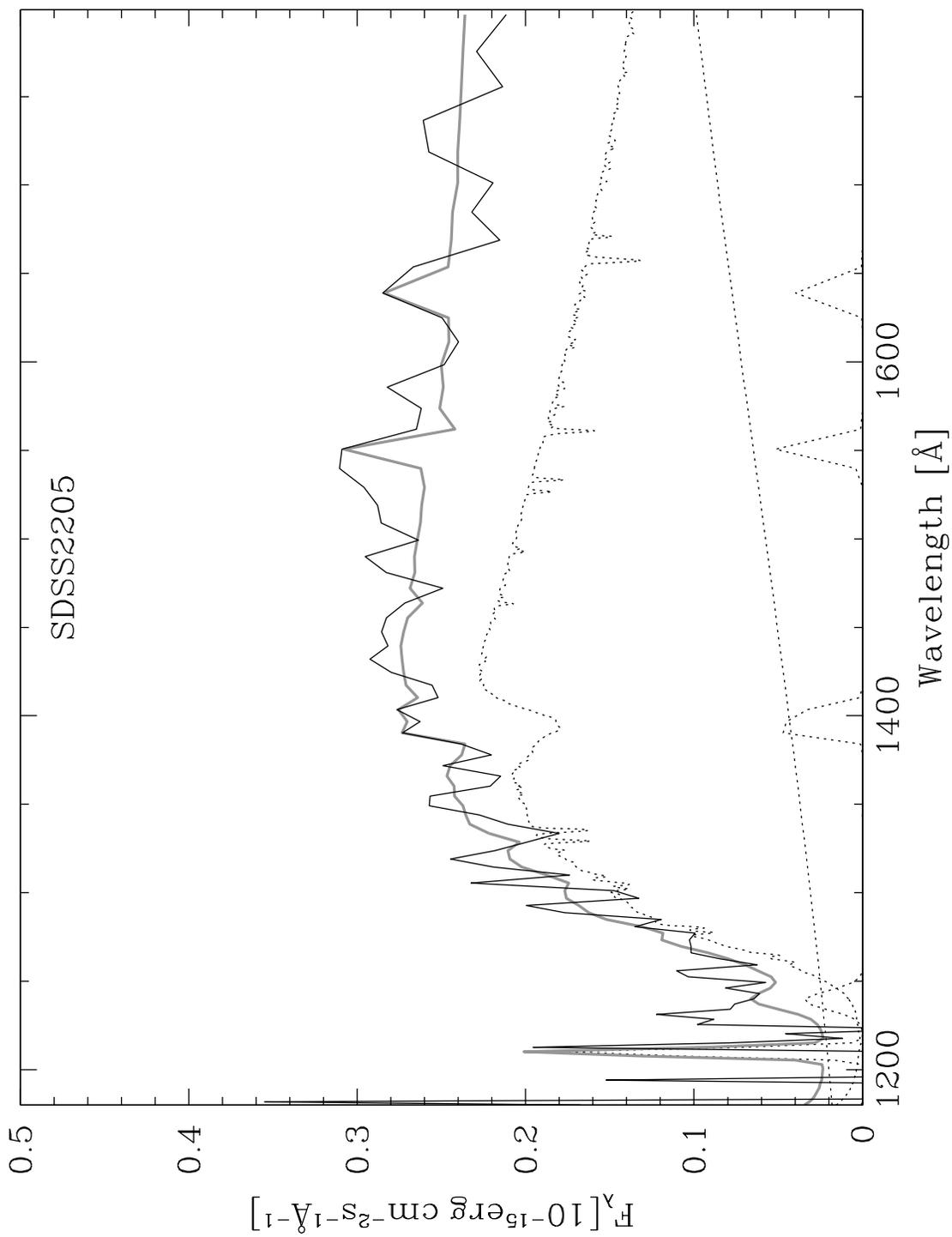}
\caption{Best fit of SDSSJ2205 with a 15,000K white dwarf at a distance of 810 pc
and a 11,000K black body.}
\end{figure}

\end{document}